\documentstyle[12pt,epsf]{article}
\newlength{\abstractwidth}
\newcommand{\onefigure}[2]{\begin{figure}[htb]
\begin{center}\leavevmode\epsfbox{#1.eps}\end{center}\caption{#2\label{#1}}

 \end{figure}}
\renewcommand{\thefootnote}{\fnsymbol{footnote}}
\renewcommand{\thanks}[1]{\footnote{#1}} 
\newcommand{\starttext}{
\setcounter{footnote}{0}
\renewcommand{\thefootnote}{\arabic{footnote}}}
\renewcommand{\theequation}{\thesection.\arabic{equation}}
\newcommand{\be}{\begin{equation}}
\newcommand{\bea}{\begin{eqnarray}}
\newcommand{\eea}{\end{eqnarray}}
\newcommand{\beq}{\begin{equation}}
\newcommand{\ee}{\end{equation}}
\newcommand{\eeq}{\end{equation}}

\newcommand{\<}{\langle}

\renewcommand{\>}{\rangle}
\def\ba{\begin{eqnarray}}
\def\ea{\end{eqnarray}}

\def\14{{1\over4}}
\def\12{{1 \over 2}}
\def\eq{&=&}

\def\d{\partial}

\def\h3{h^{3\over 2}}

\def\>{\rangle}
\def\<{\langle}
\def\sc {Schwarzschild}

\def\des{de Sitter space}

\begin{document}
\renewcommand{\theequation}{\thesection.\arabic{equation}}
\begin{titlepage}
\bigskip
\rightline{SU-ITP 00-25} \rightline{MIT-CTP-3295} 
\rightline{hep-th/0208013}

\bigskip\bigskip\bigskip\bigskip

\centerline{\Large \bf {Disturbing Implications of a Cosmological 
Constant}}

\bigskip\bigskip
\bigskip\bigskip

\centerline{\it L. Dyson$^{a,b}$, M. Kleban$^a$, L. Susskind$^a$  }
\medskip
\medskip
\centerline{$^a$Department of Physics} \centerline{Stanford
University} \centerline{Stanford, CA 94305-4060}
\medskip
\medskip
\centerline{$^b$Center for Theoretical Physics} \centerline{Department of Physics}\centerline{Massachusetts Institute of Technology} \centerline{Cambridge, MA 02139}
\medskip
\medskip

\bigskip\bigskip
\begin{abstract}
In this paper we consider the implications of a cosmological
constant for the evolution of the universe, under a set of
assumptions motivated by the holographic and horizon
complementarity principles. We discuss the ``causal patch" description
of spacetime required by this framework,
and present some simple examples of cosmologies described this way.
We argue that these assumptions inevitably lead to very deep
paradoxes, which seem to require major revisions of our usual
assumptions.
\medskip
\noindent
\end{abstract}

\end{titlepage}
\starttext \baselineskip=18pt \setcounter{footnote}{0}

\setcounter{equation}{0}
\section{Recurrent Cosmology }
As emphasized by Penrose many years ago, cosmology can only make
sense if the world started in a state of exceptionally low entropy.
The low entropy starting point is the ultimate reason that the universe
has an
arrow of time, without which the second law would not make sense.
However, there
is no universally accepted explanation of how the universe got
into such a special state. In this paper we would like to sharpen
the question by making two assumptions which we feel are well
motivated from observation and recent theory. Far from providing
a solution to the problem, we will be led to a disturbing
crisis.

Present cosmological evidence points to an inflationary beginning
and an accelerated de Sitter end. Most cosmologists accept these
assumptions, but there are still major unresolved debates concerning them.
For
example, there is no consensus about initial conditions. Neither
string theory  nor quantum gravity  provide a consistent starting
point for a discussion of the initial singularity or why the entropy
 of the initial state is so low.
High scale inflation postulates an initial de Sitter starting point with
Hubble constant roughly $10^{-5}$ times the Planck mass. This implies
an initial holographic entropy of about $10^{10}$, which is
extremely
small by comparison with today's visible entropy. Some unknown
agent initially started the inflaton high up on its potential, and
the rest is history.

 Another
problem involves so-called transplanckian modes. The quantum
fluctuations which seed the density perturbations at the end of
inflation appear to have originated from modes of exponentially
short wave length. This of course conflicts with everything we
have learned about quantum gravity from string theory.  The same
 problem occurs when studying black holes.  In the naive free field
 theory of Hawking
 radiation, late photons appear to come from exponentially
small wavelength transplanckian modes \cite{uglum}. We now know that
this is an artifact
of trying to describe the complex interacting degrees of freedom
of the horizon by quantum field theory defined on both the
interior and exterior of the black hole. A consistent approach
based on black hole complementarity \cite{stretch} describes the black
hole in
terms of  strongly interacting degrees of freedom of Planckian or string
scale, and restricts attention to the portion of the space--time
outside the horizon.

The late time features of  a universe with a cosmological
constant are also not well understood. The
conventional view
is that the universe will end in a de Sitter
phase with all matter being infinitely diluted by exponential
expansion. All comoving points of space fall out of causal contact with
one another.
The existence of a future event horizon
implies that the objects that string theory normally
calculates, such as S--Matrix elements, have no meaning \cite{quint}.
In addition, there are questions of stability of de Sitter space which
have been repeatedly raised in the past \cite{tsamis}. The apparent
instabilities are due to infrared quantum fluctuations which seem to be
out of
control. Thus the final state is also problematic.

In our opinion both the transplanckian and the late time   problems have
a common origin. They
 occur because we try to build a quantum mechanics of the
entire global spacetime--including regions which have no
operational meaning to a given observer, because they are out of
causal contact with that observer. The remedy suggested by the black hole
analogue is obvious;  restrict all
attention to a single causal patch \cite{tom,lisa,birthday}. As in
the case of black holes, the quantum description of such a region
should satisfy the usual principles of quantum mechanics
\cite{stretch}. In other words, the theory describes
a closed isolated box bounded by the observer's horizon, and  makes reference
to no other region.
Furthermore,  as in the case of black holes, the
mathematical description of this box should satisfy the
conventional principles of linear unitary quantum evolution.

Perhaps the most important conceptual lesson that we have learned
from string theory is that quantum gravity is a special case of
quantum mechanics. Thus far every nonperturbative well-defined
formulation of string theory involves a hamiltonian, and a  space
of states for it to  act on. This includes matrix theory and all
the versions of AdS/CFT. The question of this paper is whether the
usual rules apply to cosmology, and can they explain, or at least
allow, the usual low entropy starting point. In the following  we
will  assume the usual connections between quantum statistical
mechanics and thermodynamics.\footnote{After completion of this work,
we received \cite{dongsu}, which considered a similar scenario.}

These assumptions--together with the existence of a final
cosmological constant--imply that the universe is eternal but
finite. Strictly speaking, by finite we mean that the entropy of
the observable universe is bounded, but we can loosely interpret
this as saying the system is finite in extent. On the average it
is in a steady state of thermal equilibrium.  This is a very weak
assumption, because almost any  large but finite system, left to itself for a
long enough time, will equilibrate (unless it is integrable) \cite{mark}.
However, intermittent fluctuations
occur which temporarily disturb the equilibrium. It is during the
return to equilibrium that interesting events and objects form\footnote{
After completion of this work, we became aware of \cite{vilenkin}, in 
which
the authors consider a related scenario where the inflaton can repeatedly
tunnel to a false vacuum, which can result in an infinite cycle of 
new inflationary periods.}.

The essential point can be illustrated with an analogy. Instead of
the universe, let's consider a sealed box full of gas molecules.
Start with a particular low entropy initial condition with all the
molecules in a very small volume in one corner of the box. The
molecules are so dense that they form a fluid. When released the
molecules flow out from the corner and eventually fill the box
uniformly with gas. For some time the system  is far from
equilibrium. During this time, the second law insures that the
entropy is increasing and interesting things can happen. For
example, complex ``dissipative structures" such as eddy flows,
vortices, or even life can form. Eventually the system reaches
equilibrium, and all structures disappear. The system dies an
entropy death. This is the classical hydrodynamic  description of
the evolution of a ``universe". But this description is only
correct for  time intervals which are not too long.

Let $S$ be the final thermodynamic entropy of the gas. Then on
time scales of order \be T_r \sim \exp{S} \label{trec} \ee the
system will undergo Poincare recurrences \footnote{ See the appendix
for a discussion on the quantum Poincare recurrence theorem.}
\cite{lisa,birthday}. Such a recurrence can bring all the
particles back into the corner of the room. On such long time
scales the second law of thermodynamics does not prevent rare
events, which effectively reverse the direction of entropy change.
Obviously, the recurrence allows the entire process of cosmology
to begin again, although with a slightly different initial
condition. What is more, the sequence of recurrences will stretch
into the infinite past and future. The question  then is whether
the origin of the universe can be a naturally occurring
fluctuation, or must it be due to an external agent which starts
the system out in a specific low entropy state?  We will discuss
this in greater detail in Section 6.

\setcounter{equation}{0}
\section{The AdS Black Hole as a Cosmology }

It has been widely suggested that the study of cosmology has close
similarity with the study of the interior of black holes. In this
paper we will outline a framework for cosmology which is
partly inspired by the analysis of AdS/\sc \ black holes \cite{juan}.
This system, although unrealistic as a cosmology, is extremely well
understood through the AdS/CFT duality.

The relevant static metric, suitably normalized, is
\be ds^2=
\left(r^2+1 - {b_{D} \over r^{D-3}}\right)dt^2 - \left(r^2+1- {b_D \over
r^{D-3}}\right)^{-1}dr^2-r^2 d\Omega^2,
\label{adshole}
\ee
where $d\Omega^2$ represents the metric of a unit $D-2$ sphere
and $b_D$ depends on the $D$-dimensional Newton constant
and the volume of the $D-2$ sphere.
The horizon is located at $r_+$, the largest solution to
the equation
\be
r^2+1 - {b_D \over r^{D-3}} =\ 0.
\ee
The boundary is at $r=\infty$. In general this geometry will be
accompanied by a compact factor such as a 5-sphere for the case of
${\rm AdS}_5$. We will refer to the boundary as the ``observer."
Observations by the observer are known to be completely described
by a unitary holographic quantum mechanics with no loss of
information. Furthermore the black hole from the observer's
viewpoint is just the thermal state of the holographic dual.

\onefigure{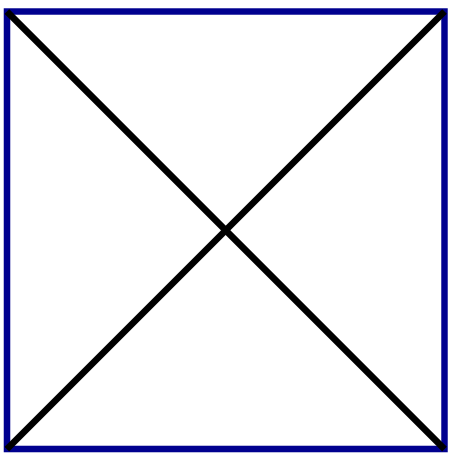}{An eternal AdS Schwarzshild black hole, or de Sitter
space.  The horizontal boundaries are space-like singularities in the case
of the AdS black hole, and the infinitely inflated past and future in the
case of de Sitter.}

Now consider the classical global geometry represented by a Penrose
diagram as shown in figure (\ref{fig1}).
It is evident from the diagram
that the classical geometry can be interpreted as a cosmology.
To the observer at the boundary, the entire universe can be
described by the triangle in figure(\ref{fig2})
\onefigure{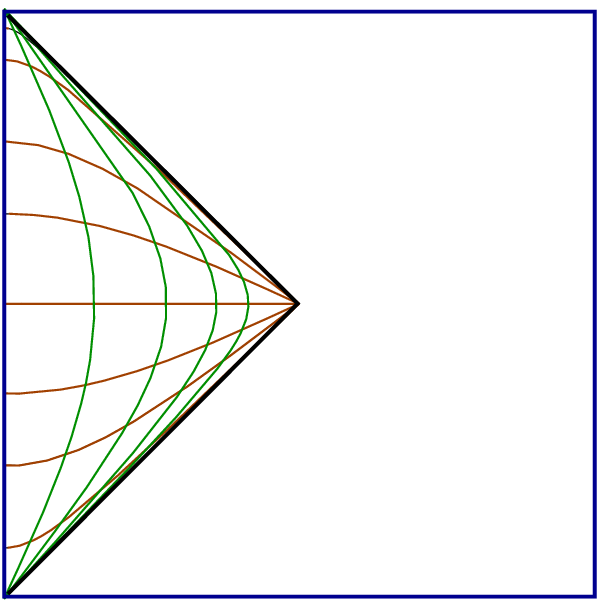}{The causal patch for an observer at the boundary of
an AdS Schwarzschild black hole, or in de Sitter space.}
which we can call the
causal patch. It describes the exterior of the black hole, defined
by $r>r_+$. This region is described by a thermal density matrix in
the exact holographic quantum theory. From the bulk viewpoint the
Bekenstein entropy is the entropy of the causal patch.

On the other hand, the global geometry is more or less similar to
closed FRW cosmologies.  It  starts with a big bang singularity
and ends with a final crunch. An observer at the spatial center of
the diagram will experience the crunch in a finite time. One
peculiar feature is that the rapid time dependence near the early
singularity is far from adiabatic. Under these kind of
circumstances there should be large amounts of entropy production.
On the other hand the geometry is obviously time symmetric. This
is very unnatural and presumably means that the universe starts in
very special, extremely fine tuned  state. A rough analogy would
be a history in which all the particles in a box start in one
corner and after a given long time return to the corner. Such
trajectories will exist but they are extremely unstable. Moving
one particle a little bit will change the outcome to a more
thermodynamically sensible final state.

A more reasonable cosmology can result if we started with a far
from equilibrium configuration. For example, an initially empty
AdS space of zero entropy could be excited by an infalling shell
of energy originating at the boundary. From the dual conformal
field theory point of view, this corresponds to varying the
hamiltonian in a time dependent way. For example, by varying the
gauge theory coupling a coherent pulse of dilaton wave can be
injected into the bulk.
The ability to modify the action in this way and still have a
solution to the supergravity equations of motion is of course
connected with the existence of an ultraviolet boundary.

The dilaton pulse will create some thermal energy.  If it has
enough energy, an AdS  black hole will form. The observer in the
bulk will start life in empty AdS with zero entropy. He will then
see some sudden time dependence as the dilaton wave passes.
Afterwards the bulk will become excited with heat and particles
and possibly dissipative structures. Eventually the bulk observer
will be destroyed at the singularity. This cosmology is shown in
figure (\ref{fig3}).
\onefigure{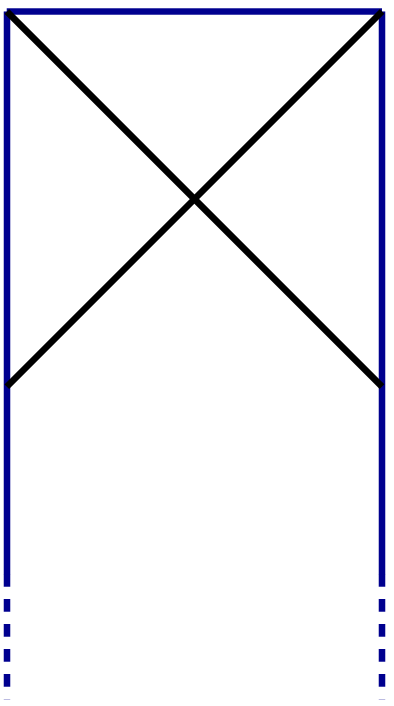}{An AdS black hole created by an infalling shell of energy
originating at the boundary.}

From the CFT side, the black hole formation is just the classical
gravitational description of the approach to thermal equilibrium.

As in the case of the box of molecules, this description does not
make sense for time intervals of order $\exp{S}$. Poincare
recurrences will lead to intermittent rare fluctuations in which
the energy of the black hole reassembles itself into an energy
shell, which goes out and then reflects back. In other words, the
formation of the black hole will endlessly repeat itself at
intervals of order $T_r \sim
\exp{S}$. Each recurrence will be different in its microscopic
details from the previous.

One important fact is that these recurrences cannot be described
by classical general relativity. It is as though classical
relativity--with its thermodynamic-like laws of black hole
mechanics--is a coarse grained version of a more detailed theory.
This raises the obvious question of whether the initial conditions
of the universe could have originated from a Poincare recurrence;
that is, a large thermal fluctuation to a very low entropy
configuration.

In the recurrent view of cosmology the second law of
thermodynamics and the arrow of time would have an unusual
significance. In fact they are not laws at all. What is true is
that interesting events, such as life, can only occur during the
brief out-of-equilibrium periods while the system is returning to
equilibrium. In other words, the second law and the existence of a
time arrow are anthropic. They are part of the environmental
conditions needed for life.

\setcounter{equation}{0}
\section{Thinking Inside The Box }

For reasons that have been explained in many places, if we want to
apply the ordinary principles of unitary quantum mechanics to a
system with horizons, we need to restrict attention to the region
on the observer's side of the horizon. For example, the
triangle of figure (\ref{fig2}) is the the region appropriate to an
observer on the left-hand boundary of the space. We call such a
region the $observer's\ \ causal\ \ patch$.

The space that we will consider next is de Sitter space. The
causal structure of \des \ is described by the same Penrose
diagram as shown in figures (\ref{fig1}) and (\ref{fig2}), but with
different
interpretations. The horizontal boundaries which previously
described singularities now represent the infinitely inflated past
and future (IIP and IIF). The vertical edges are no longer true
boundaries, as they were in AdS, but represent observers at
opposite poles of a global spatial sphere. The triangle of figure
(\ref{fig2}) is
now the causal patch of one observer. One point in common with the
AdS black hole is that the causal patch is described
thermodynamically. The entropy usually ascribed to \des \ is that
of the causal patch. As in the black hole case, it is the logarithm
of the number of effective quantum states of the thermal ensemble.
More exactly it is ${\rm Tr} \rho \log{\rho}$, where $\rho$ is the
density matrix describing the causal patch.


The causal patch can be described by a static metric
\be
ds^2=R^2 \left[(1-r^2)dt^2 - (1-r^2)^{-1}dr^2-r^2
d\Omega^2\right],
\label{statmet}
\ee
where $R$ is the inverse Hubble scale. The radial coordinate runs
from $r=0$ at the location of the observer to $r=1$ at the
horizon. The horizon  is the boundary of a kind of box or cavity
with the observer at the center. Because the boundary of the box
is a horizon, the box has a finite temperature and entropy:
\bea
T \eq {1\over 2\pi R} \cr \cr
S \eq {4 \pi R^2 \over 4G}
\label{TS}.
\eea

We are  interested not only in the classical description of
``dead" de Sitter space, but also the non-equilibrium fluctuations
that can lead to an interesting universe. In order to illustrate
the main ideas, we will study a simplified model in which the
universe will be taken to be $(2+1)$ dimensional, so that $\Omega$
in (\ref{statmet}) is replaced by a single angle $\theta$.
Furthermore, we will restrict discussion to configurations with
rotational symmetry and no angular motion.

Our first task is to generalize the metric in (\ref{statmet}) to
non-static configurations. In particular, we want to introduce
coordinates which cover only the region in full causal contact
with the observer at $r=0$. By this we mean the region which can
both receive signals from, and transmit signals to, the observer.
To accomplish this, it is convenient to first change the radial
coordinate in (\ref{statmet}) so that the metric has the form
\be
ds^2 = e^v(dt^2-d\rho^2) -Z^2 d\theta^2.
\label{tort}
\ee
The transformation for pure de Sitter space is given by:
\bea
r \eq \tanh{{\rho \over R}}
\cr \cr e^v \eq {1\over \cosh^2{{\rho \over R}}}
\cr \cr Z \eq R \tanh{{\rho \over R}}.
\label{rho}
\eea
The entire causal patch is covered by the region
$\rho \geq 0, \ \ -\infty <t< +\infty .$

Now let us introduce a gauge fixing procedure which leads to
coordinates of this type, for arbitrary geometries of the form we
are considering. In $(2+1 )$ dimensions we can choose three gauge
fixing conditions. We choose these to be
\bea
g_{tt} \eq -g_{rr} \cr
g_{tr} \eq 0 \cr
g_{r \theta}\eq 0
\label{fix}.
\eea
In addition, given the conditions of rotational symmetry and no
angular velocity, the  component $g_{t \theta}$ vanishes. Thus the
metric has the form (\ref{tort}). However we have not used all the
gauge fixing freedom at our disposal. Let us define coordinates
$t^{(\pm)} =t \pm \rho$. Conformal transformations of the form
$X^{\pm}=X^{\pm}(t^{\pm}) $ preserve the form of the metric. This
means we can still choose two functions, each of one variable.
Thus we make the additional gauge choices
\bea
e^v|_{\rho=0} \eq 1 \cr \cr
Z|_{\rho=0} \eq 0.
\eea
This completely fixes the gauge, and results in coordinates which
exactly cover the causal patch of the observer at $\rho =0$
\footnote{We have made the assumption that the proper time along
the observer's trajectory goes to infinity in the remote past and
the remote future; i.e., the trajectory does not encounter a
singularity in finite proper time.}.  

We could also choose to fix the guage by requiring regularity in the
coordinates at the horizon, rather than at the origin.  
Once the observer's location in the infinite past and future are 
chosen, the causal patch horizon is determined, and therefore such a 
gauge choice is manifestly independent of the observers trajectory.
Furthermore, it is the appropriate choice if one wishes to define a 
fluctuating energy.  As we will see, fixing the gauge at the
origin makes the energy independent of the de Sitter radius. 

In general, $e^v$ and
$Z$ will be time dependent. An important point is that at
the location of the observer, the coordinate $t$ is precisely the
proper time of a clock carried by the observer. We will call this
gauge the observocentric-gauge, or O-gauge. The coordinates $t,\rho$
can easily be constructed for some special cases of interest. In
particular, consider a closed FRW metric 
\be
ds^2 = dT^2 - a(T)^2 
\left[d\theta^2 + \sin^2(\theta) d\phi^2 \right], 
\ee
which begins in the
remote past as a de Sitter space with radius $R'$ and ends in the
remote future as a de Sitter space with radius $R$. An
inflationary universe with a final nonzero cosmological constant
is an example. The scale factor $a$ is a function only of FRW
time $T$.  To construct the O-gauge
coordinates $\rho$ and $t$ at an arbitrary spacetime point
$p$, consider past and
future light rays originating at  $p$ and propagating to
the observer. These intersect the observer's trajectory at 
past and future FRW times $T_{\pm}$.
The O-gauge coordinates of the point $p$ are then
\bea
2t \eq T_+ + T_-
\cr 2\rho \eq T_+ - T_- .
\label{coord}
\eea
It is also easy to find the values of $e^v$ and $Z$.  Define the 
function $\tau(x) = \int^x dT / a(T)$, its inverse $\tau^{-1}$, and $t_\pm 
= t \pm \rho$.  Then
\bea
e^v \eq {a^2 \left( \tau^{-1}\left( {1 \over 2} \left[ \tau(t_+) + 
\tau(t_-)\right]\right) \right) 
\over a(t_+)  a(t_-) } \cr \cr 
Z \eq a^2 \left( \tau^{-1}\left( { \tau(t_+) +
\tau(t_-) \over 2 } \right) \right) 
\sin^2 \left( { \tau(t_+) - \tau(t_-) \over 2} \right) .
\label{FRW}
\eea

One can easily represent spatially flat inflationary universes in a
similar manner; in general, the O-gauge coordinates are constructed so as
to exactly cover the observer's causal patch.  The time variable at $\rho
= 0$ (the observer) is always proper time. The construction can be
generalized to include non-rotationally invariant geometries, in arbitrary
dimension. Although we will not discuss this in detail here, one would
construct the causal patch coordinates simply by considering the
intersection of the future and past lightcones of two points $\tau_{\pm}$
along the trajectory. These will intersect along a $d-2$ sphere of
constant $t$ and $\rho$, as given by \ref{coord}.

\begin{figure}[htb]
\begin{flushleft}
\leavevmode\epsfbox{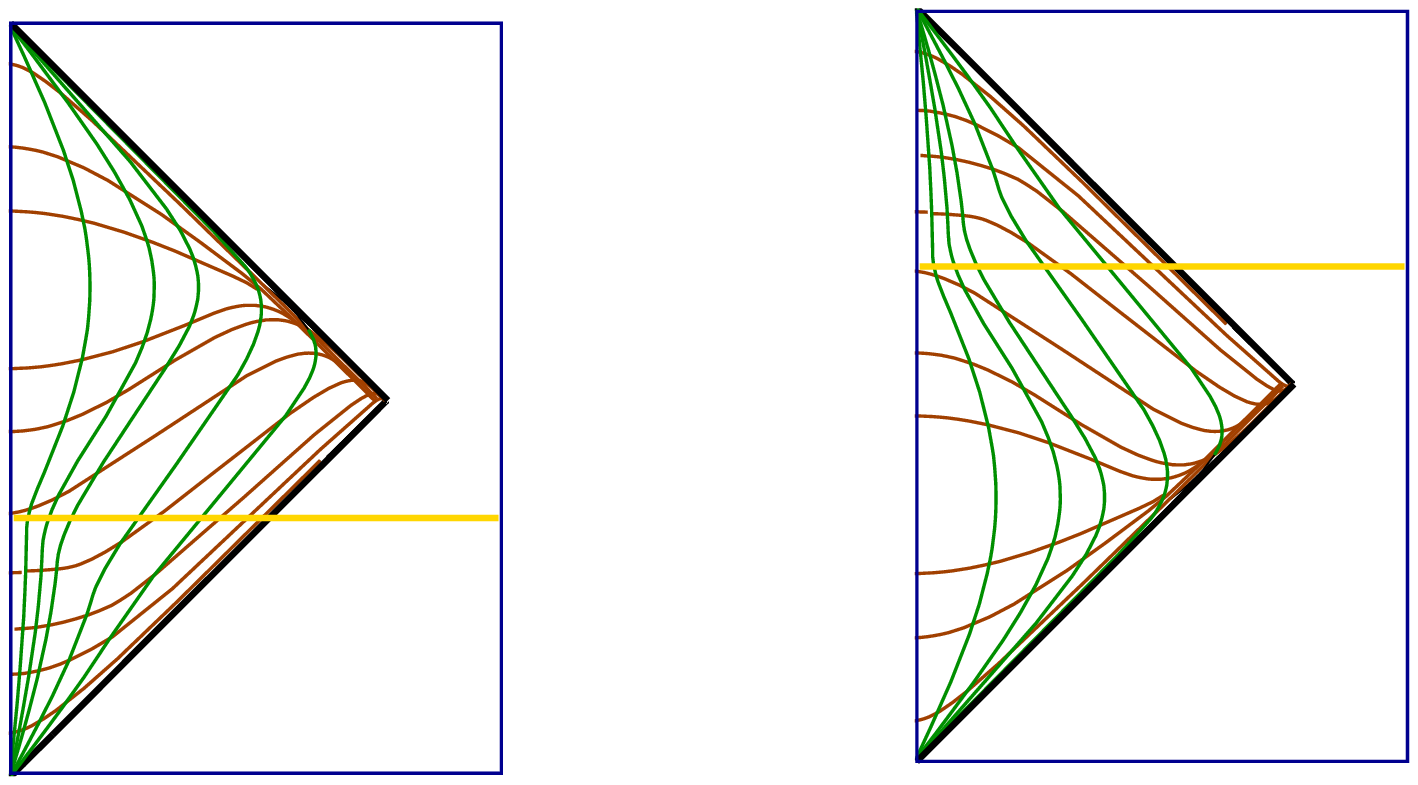}
\caption{The causal patch for two types of global time
transitions from one de Sitter space to another. \label{fig4}}
\end{flushleft}
\end{figure}

There are two cases that can be distinguished by the ``height" of
the Penrose diagram. If the height is less than twice the width, as
in the two spacetimes of figure (\ref{fig4}),
then the surfaces $t=\pm \infty$ intersect within the
diagram at a Rindler type horizon, just as they would in de Sitter
space. If, however, the diagram is taller than twice its width,
as in figure (\ref{fig5}),
\onefigure{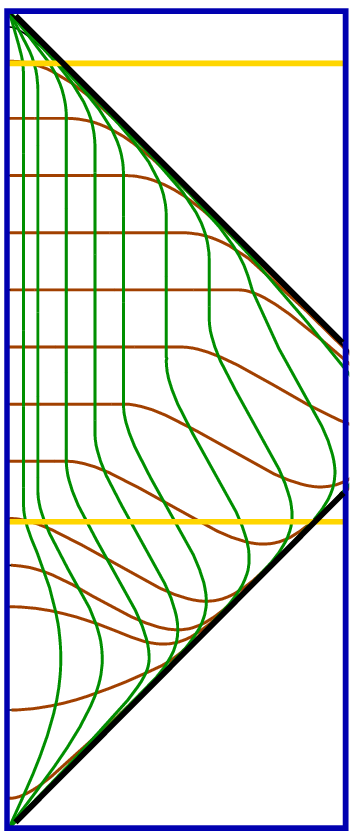}{A transition from a de Sitter space with a
small radius of curvature to a matter dominated phase, followed
by a transition to a large de Sitter.  In this example, the
horizon is pushed all the way around to the south pole
of the sphere, and
vanishes for intermediate times.}
the two surfaces terminate at the antipodal point before
intersecting. In this case the spatial slices $t=const$ form
closed spatial spheres. In either case, the surfaces $t=\pm \infty$
are the past and future horizons of the observer.

It is also convenient to stretch the horizon in a way familiar
from black hole physics. One way to do this is to define
a time-like surface as follows:  begin either at the intersection
of the past and future horizons, or at the point where the spatial
surfaces meet the antipode. From that point work back  (toward
$\rho=0)$ along the surfaces of constant $t$ a fixed proper
distance of order the Planck or string scale. The resulting
position is the stretched horizon.

For the case in which the geometry begins and ends in de Sitter
phases, the stretched horizons at early and late times are the
same as we would expect for the corresponding eternal de Sitter
spaces. This implies that the horizon entropy within the causal
patch in the far future and past is bounded in the ordinary way.
In particular, no matter how many e-foldings of inflation occur
before the spacetime exits to its final configuration, the entropy
in the final patch must still be bounded by the area of the horizon,
as determined by the cosmological constant.  This bound can easily
be shown to be equivalent to the covariant entropy bound as proposed
in \cite{bousso}.

\setcounter{equation}{0}
\section{Some Examples }
Here are some examples of thinking inside the box. Let's begin
with a minimally coupled scalar field coupled to gravity. The
action is given by
\be
I= \int d\rho d\theta dt \sqrt{-g} \left( R + g^{\mu \nu}
\partial_{\mu}\phi \partial_{\nu}\phi -V(\phi)\right).
\label{action}
\ee
In the O--gauge the action is
\be
I= \int d\rho d\theta dt
\left( \dot{Z} \dot{v}-Z'v'
+Z ( \dot{\phi}^2-(\phi')^2)
-e^vZ V(\phi)
\right).
\label{oaction}
\ee

It is straightforward to construct a Hamiltonian from this action. Its
value is given by \be H= \int d\rho d\theta \left(\dot{Z} \dot{v}+Z'v' +Z
( \dot{\phi}^2+(\phi')^2) +e^vZ V(\phi) \right). \label{ham} \ee
Classically the value of $H$ is constrained by the Einstein equations. The
time--time component is given by 
\be \dot{Z} \dot{v}+Z'v' +Z (
\dot{\phi}^2+(\phi')^2 ) +e^vZ V(\phi)- 2 Z'' = 0. \label{ein} \ee By
integrating this equation over all space we find \be H-2 \d_{\rho}
Z|^{\rho=\infty}_{\rho =0}=0. \label{intein} 
\ee 
The boundary contribution
at $\rho = \infty$ vanishes in general, and for geometries which are
smooth at the origin, $\d_{\rho}Z=1$\footnote{This ``boundary term at the
origin" is special to 2+1 dimensions, and vanishes in higher $d$.}. Thus
the constraint requires $H=2$. As mentioned above, this is completely
independent of the de Sitter curvature, and of any non-singular matter
content in the spacetime (had we chosen to fix the guage at the horizon,
the energy would depend on the far past and future values of the de Sitter
curvature).  Therefore, complicated dynamical processes such as the
evolution from a small de Sitter space to a large one, with some periods
of matter or radiation domination in between, can all occur at constant
energy\footnote{This definition of the energy is identical to the one
proposed in \cite{horowitzhawking}, with the spatial slicing being
surfaces of constant $t$, and the causal horizon taken to be the
boundary.}.
We will illustrate this point with some specific examples below.

Solutions can easily be generated from FRW solutions transformed
into the O--gauge. As an example, we take the potential $V$ to be
a constant plus a small step-like function:
\be
V(\phi)=V-c \tanh{L\phi},
\label{vee}
\ee
with $c<<V$. There are solutions which in FRW coordinates roll
from negative to positive $\phi$, going over the step at some
fairly localized time. This time can be either during the
contracting or the expanding phase of the FRW description. Roughly
speaking there is a small decrease of the cosmological constant at
this time. The change is small so the back reaction on the
geometry is also small.

In O--coordinates, the behavior depends on whether the transition
occurs during FRW contraction or expansion. In the former case, a
domain wall or wave of decreased cosmological constant originates
at the horizon ($\rho = \infty$) and propagates inward toward the
observer. Energy is released, and since the final entropy of the
horizon increases, we may describe the domain wall as producing
entropy which falls to the horizon.

In the latter case that the transition occurs during expansion,
the O--description is quite different. In this case the domain
wall originates at the observer and falls toward the horizon.
Again the final effect is to increase the horizon entropy.

The back reaction on the geometry is small, but it results in
making the Penrose diagram slightly taller. The effect is to push
the horizon point slightly toward the antipodal point as in figure
(\ref{fig4}). To understand this effect, first consider the causal patch
in
pure \des . The patch is perfectly static and consists of a
hemisphere (southern). The observer is at the south pole. The
 horizon is at the equator, and stretching it  displaces it slightly 
south.



In the time dependent solution the stretched horizon begins in the
past at its static location but then temporarily moves away from
the observer toward the north pole. Eventually it returns to very
near the equator.

One can also study the more extreme case in which there is a very
large change in the cosmological constant, as in a realistic
inflationary theory. Typically, in this case, the horizon
intersection moves far toward the north pole.  In some cases it
may even reach the pole, so that the causal patch may temporarily
become a closed sphere, with the horizon disappearing for a time.
We illustrate an example of this in figure (\ref{fig5}).

As another example of thinking in the box, consider a mass
distribution in de Sitter space.  For simplicity, we will work
with a spherically symmetric matter distribution localized along a
thin ring. The solution outside of the ring will have a deficit in
the range of the angular variable. The solution can be written
\be
ds^2 = R^2\left[(1 - r^2)dt^2 - (1 - r^2)^{-1} dr^2 - \alpha^2 r^2 d
\theta^2\right]
\hskip 30 pt (r < r_0)\ .
\ee
where $r_0$ is the position of the ring, $\theta \simeq \theta + 2 \pi$
and
$0 < \alpha \leq 1$ determines the
deficit angle.
 Note that the area of the horizon is $2 \pi \alpha \leq 2 \pi$.  We
interpret this as meaning that some of the degrees of freedom
making up the horizon are ``used up" in forming the ring. If the
size of the ring is much smaller than the de Sitter radius, we can
think of it as an object which could exist in the corresponding
theory without a cosmological constant. In that case it is natural
to identify the deficit angle with the mass. Evidently then, the
mass is proportional to the amount of entropy used up in making
the object. More precisely, the number of degrees of freedom
borrowed from the stretched horizon is the product of the mass and
the \des \ radius.

Inside the ring, the solution is pure de Sitter space with no deficit
angle.  The metric is
\be
ds^2 = R^2\left[(1 - r'^2)dt '^2 - (1 - r'^2)^{-1} dr'^2 - r'^2 d
\theta^2\right]
\hskip 30 pt (r' < r'_0)\ .
 \ee
The condition that the metric be continuous across $r
= r_0$ requires $r' = \beta r + r_0(\alpha - \beta)$ and $t\ ' =
\beta^{-1}
t$,
with $\beta^2 = (1 - \alpha^2 r_0^2)(1 - r_0^2)^{-1}$ and $r'_0 = \alpha
r_0$.

\onefigure{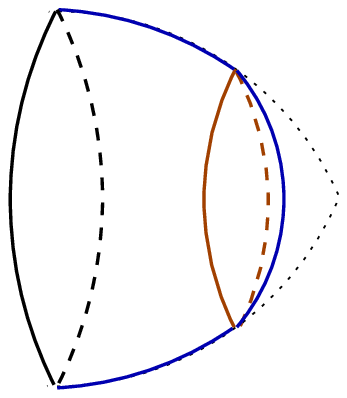}{Cigar diagram for a ring mass in de Sitter space.  The
left-hand boundary is the horizon.  The dotted line indicates the shape
for the case of a point mass at the origin.}

One can compute the stress-energy required for this solution to exist.
The contribution
from the ring is
\be
\sqrt{g} T^t_t =  \sqrt{1 - r_0^2} \left( \alpha \sqrt{1 - r_0^2} -
\sqrt{1 - \alpha^2 r_0^2} \right)
\delta(r - r_0),
\ee
and
\be
\sqrt{g} T^\theta_\theta =  \alpha r_0^2 \left( 1 - \alpha \sqrt{(1 -
r_0^2) (1 - \alpha^2 r_0^2)^{-1}} \right)
\delta(r - r_0).
\ee

This corresponds to a positive energy density in the ring (for
$\alpha > 0$), with some tension to hold it static. This solution
is unstable to non-spherically
symmetric perturbations, but this is not important for our purposes.

By making the appropriate coordinate transformations (\ref{rho}),
we can easily express the above solution in O-gauge (\ref{tort}).
The energy of the configuration then is given by (\ref{intein}),
$H = 2$.  This solution makes manifest that the energy is
independent of the matter content of the space. The time-time
component of Einstein's equations always contains second
derivative terms, and we define energy by the boundary terms
corresponding to these.  In $2+1$ dimensions there are two
possible contributions:  at the origin and at the horizon. As we
have already discussed, the contribution at the horizon generally
vanishes. Very close
to the origin the metric looks flat, and $Z \sim \rho$ as long as
the mass distribution is not localized there. Since $\partial_\rho
Z = 1$, the energy is independent of the matter content of the
space.

As the simple examples above show, a fluctuation which removes
some degrees of freedom from the horizon (thereby decreasing its
entropy), and deposits them in some configuration in the bulk of
the space--perhaps by increasing the vacuum energy, or in the form of
matter or radiation--is an energy conserving process.  The observable history
of the universe is one such fluctuation, occurring at fixed energy
in some background de Sitter space.

\setcounter{equation}{0}
\section{Complementarity and its Implications }

At this point we will add one more assumption, which is motivated  by the
success of string theory in describing black holes using the
standard principles of quantum mechanics. The additional
ingredient is the analog of black hole complementarity. In more
general contexts it can be called Horizon Complementarity. We
assume that the physics within the causal patch is exactly
described by a dual quantum system that includes a Hamiltonian $H$
and a space of quantum states. Furthermore the state of the world
within the causal patch is density matrix $\exp{(-\beta H)}$ with
the appropriate temperature. The assumption is motivated by what
we now know about other quantum gravity systems such as the AdS
black hole \footnote{The conjectured quantum dual that we are
discussing in this paper should not be confused with the attempts
to define a dS/CFT correspondence. The quantum dual in this paper
is not identified with the space--like boundary of
\des. }.

We know a few things about the dual quantum system. From the fact
that the entropy is finite, it follows that the energy levels are
discrete and that the number of levels below any given energy is
finite. That is all we really need, but there is also reason to
believe that there may be an upper bound on the energy levels
\cite{tom}. One way to argue this point is to observe that in
other holographic dual theories, the UV sector of the theory is
identified with the timelike  boundary of the space. In the de
Sitter case, the closest analog of a timelike boundary is the line
$r=0$, suggesting that the high-energy degrees of freedom are
sparse or non-existent above a certain level.  By contrast, the
low energy degrees of freedom are associated with the horizon of
the causal patch. Since the area of the horizon at $r=1$ is larger
than at any other radius, we expect the system to be very rich in
low energy degrees of freedom.  If this is correct, the quantum
dual could be a system with a finite dimensional Hilbert space as
advocated by Banks \cite{banksb} and Fischler \cite{fisch}.

The implications of
Horizon Complementarity are profound, and may lead
to new  cosmological questions and puzzles. As we shall see, it implies that
the universe has neither a beginning nor and end. Instead, it
intermittently recycles itself, but in a way that is very
different from so called cyclic universes.

For our present purposes, Horizon Complementarity will be taken to
mean that physics in the causal patch can be described in terms of
a closed or isolated system described by conventional quantum
mechanics. Furthermore we assume that the maximum entropy that the
system can ever have is given by the entropy of a de Sitter space,
whose size is determined by the cosmological constant. By the
cosmological constant, we mean the true minimum value of the
vacuum energy, not the value during early inflation. For
illustrative purposes we can take the cosmological constant to be
the current value of the dark energy that appears to be
accelerating the expansion of the universe today. Horizon
Complementarity then says the universe is described by any
observer as a closed and finite system.

The implication of such a description, as we have suggested in
Section (1), is that Poincare recurrences are inevitable. Starting
in a high entropy, ``dead" configuration, if we wait long enough,
a fluctuation will eventually occur in which the inflaton will
wander up to the top of its potential, thus starting a cycle  of
inflation, re--heating, conventional cosmology and heat death. The
frequency of such events is very low. The typical time for a
fluctuation to occur is of order
\be
T_r \sim \exp{( S-S')}
\label{tfluc}
\ee
where $S$ is the equilibrium entropy and $S'$ is the entropy of
the fluctuation. The fluctuations we have in mind correspond to
early inflationary eras during which the entropy is probably of
order $10^{10}$, while the equilibrium entropy is of order
$10^{120}$. Thus
\be
T_r \sim \exp{10^{120}}.
\ee
This seems like an absurdly big time between interesting events,
which by comparison last for a very short time. Nevertheless
dismissing such long times as ``unphysical" may be a symptom of
extreme temporal provincialism.

Given our assumptions, the conclusion that such Poincare recurrences of
the universe occur is unavoidable. The potential difficulty is not the
long time between them, but rather that there may be far more probable
ways of creating livable (``anthropically acceptable") environments than
those that begin high up on the inflationary potential (to height $V$).

In the next section, we will discuss the probability that a
recurrence will lead to a world which is consistent with our own
universe.  First, however, there is an alternative that we should
explore. It is suggested by the discussion of the AdS/\sc \ black
hole in section (2) where the eternal black hole was replaced by a
black hole formed from a shell of energy that originated at the
boundary. From the dual point of view the system corresponds to a
time dependent deformation of the boundary conformal field theory
which heats the system. The question is whether a similar story
makes sense for
\des ? We think the answer in no. The reason is that the causal
patch does not have the analog of an ultraviolet boundary. Because
AdS has such a boundary it is possible to modify the boundary
conditions without changing the fact that the bulk theory still
satisfies the bulk field equations. Without such a boundary there
is no way to do the same for \des . Thus, if there is a theory of
\des \ and its fluctuations we do not expect to be able to
intervene from the outside and create a particular initial
condition. In our view, the only possible origin of an
out-of-equilibrium initial state is a large fluctuation.

We will conclude this section with a few remarks about the
proposed dS/CFT duality that has been the subject of several
recent papers. The idea is to use the space-like boundary of \des
\ in place of the time-like boundary of AdS in order to formulate
a holographic theory. In particular deformations of the boundary
conditions at past infinity could allow us to introduce initial
conditions such as an inflationary era. While this is an
attractive idea, it does not seem consistent with the thermal
properties of \des . To understand why, we can try to interpret
the idea in the box. For a perturbation on the past boundary to
influence the causal patch, the signal must pass through the past
horizon of the observer. However the past horizon is just the
infinite past $t= - \infty$. In other words, it simply corresponds
to a perturbation of the box at infinitely negative time.
Obviously, any such perturbation of a finite entropy thermal
cavity will re-thermalize after a finite time.  In fact, at any
finite time, an infinite number of Poincare recurrences have
already occurred.

\section{The Misanthropic Universe?}

Certainly, given enough time and a suitable inflaton, recurrences
will eventually bring the box to a configuration that could serve
as an initial starting point for a standard inflationary theory.
The entropy of such configurations is very low. In a typical
high-scale inflationary theory, the entropy of the initial de
Sitter-like space can be as low as $10^{10}$. The entropy of the
final \des \ (assuming the observed value for the dark energy) is
of order $10^{120}$.

Inflationary starting points are very rare in time. That in itself
is not a problem. Most of the rest of time during which nothing
interesting is happening can, from an anthropic point of view,  be
thrown away. We can also throw away large fluctuations which lead
to un--livable conditions. The danger is that there are too many
possibilities which are anthropically acceptable, but not like our 
universe.

Let us consider the  entropy in observable matter in today's
universe. It is of order $10^{100}$. This means that the number of
microstates that are macroscopically indistinguishable from our
world is  $\exp{(10^{100})}$. But only $\exp{(10^{10})}$
 of these states could have evolved from the low entropy initial
 state characterizing the usual inflationary starting point. The
 overwhelming majority of states which would have evolved into a
 world very similar to ours did not start in the usual low entropy
 ensemble.

 To understand where they came from, imagine running these states
 backward in time until they thermalize in the eventual heat bath
 with entropy $10^{120}$. Among the vast number  $\sim
 \exp{(10^{120})}$ of possible initial starting points, a tiny
 fraction $ \sim \exp{(10^{100})}$ will evolve into a world like
 ours. However, all but  $ \sim \exp{(10^{10})}$ of  the corresponding trajectories
(in phase space) are extremely unstable to tiny perturbations.
Changing the state of just a few particles at the beginning of the
trajectory will lead to completely different states. Nevertheless,
there are so many more of these configurations among the
$\exp{(10^{120})}$ possible starting points that in a theory that
relies on statistical likelyhood to fix the initial conditions,
they completely dominate.

What is worse is that there are even more states which are
macroscopically different than our world but still would allow
life as we know it. As an example, consider a state in which we
leave everything undisturbed, except that we replace a small
fraction of the matter in the universe by an increase in the
amount of thermal microwave photons. In particular, we could do
this by increasing the temperature of the CMB from $2.7$ degrees
to $10$ degrees. Everything else, including the abundances of the
elements, is left the same.  Naively one might think that this is impossible; no consistent
evolution could get to this point since the extra thermal
energy in the early universe would have destroyed the fragile nuclei.
But on second thought, there must be a possible starting point which would
eventually lead to this ``impossible" state. To see this, all we have to 
do is
run the configuration backward in time. Either classically or quantum
mechanically, the reverse evolution will eventually lead back to a state which
looks entirely thermal, but which, if run forward, will lead us back to where we
began. In a theory
dependent on Poincare recurrences, there would be many
more ``events," in which the universe evolves into this modified
state. Thus it would be vastly more likely to find a world at $10$
degrees with the usual abundances than in one at $2.7$ degrees.
All of these worlds would be peculiar. The helium abundance would
be incomprehensible from the usual arguments. In all of these
worlds statistically miraculous (but not impossible) events would
be necessary to assemble and preserve the fragile nuclei that
would ordinarily be destroyed by the higher temperatures. However,
although each of the corresponding histories is extremely
unlikely, there are so many more of them than those that evolve
without ``miracles," that they would vastly dominate the livable
universes that would be created by Poincare recurrences. We are
forced to conclude that in a recurrent world like \des \   our
universe would be extraordinarily unlikely.

\begin{figure} 

\begin{center} 

\leavevmode\epsfbox{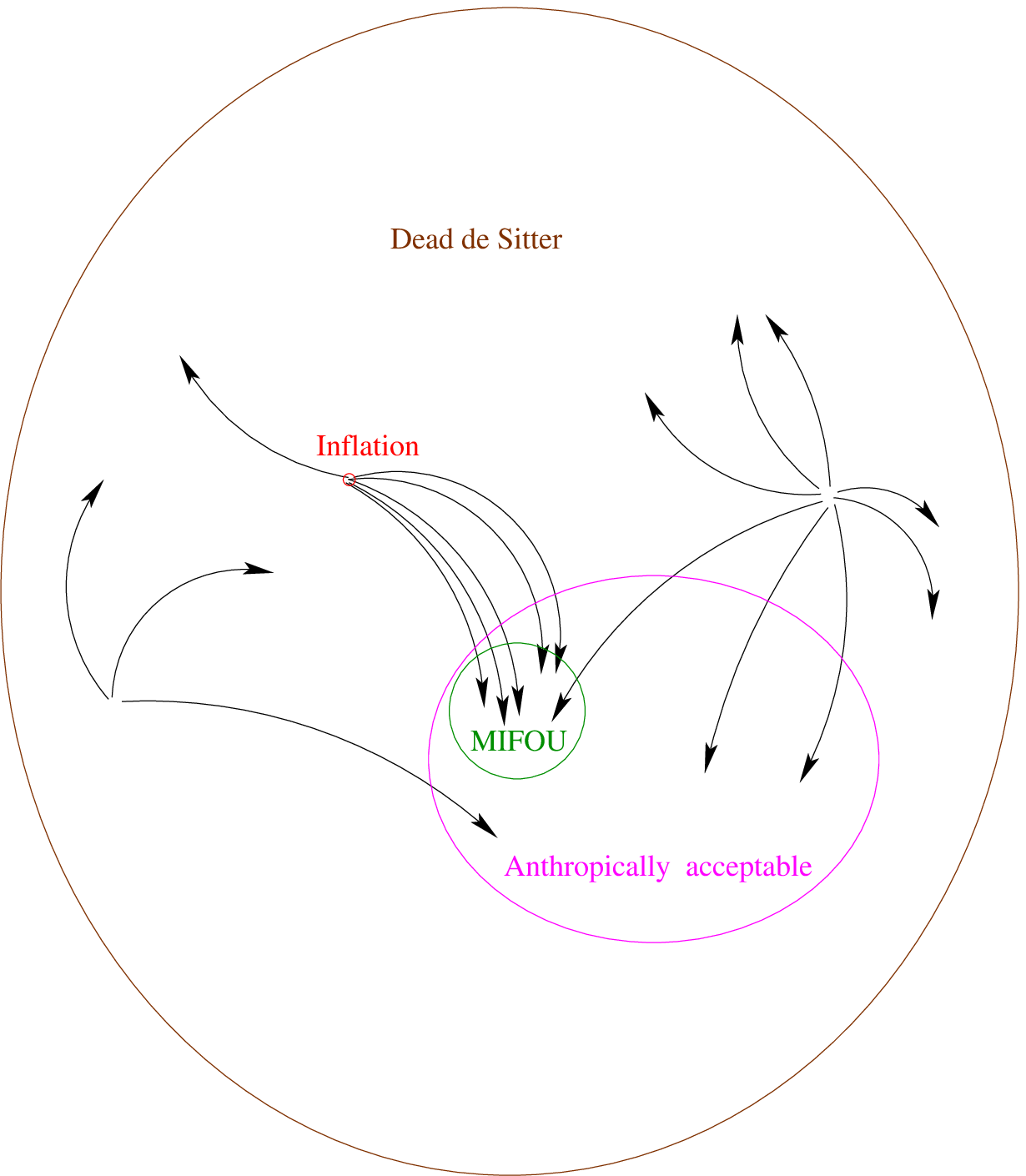} 

\caption{A schematic representation of the phase space of a universe with
a cosmological constant. The vast majority of the space consists of states
which are macroscopically ``dead de Sitter;" that is, nearly empty de 
Sitter containing only some thermal radiation.  A tiny subset of
the states are anthropically acceptable, meaning that they contain complex
structures such as stars and galaxies, and a very small subset of those
are macroscopically indistinguishable from our universe (labelled MIFOU in
the figure).  Inflationary initial conditions occupy an even smaller
fraction of the space.  Trajectories which pass through the inflationary
patch will almost always lead immediately to the MIFOU region, ``mixing"
into it in a ``porous," phase-space-area-preserving manner.  The vast
majority of the points in the MIFOU region did not come from inflation,
but rather from unstable trajectories originating in the dead region.
Finally, any trajectory in the dead region will remain there almost all of
the time, but will occasionally enter the athropically acceptable region,
and very much more rarely the MIFOU region, and almost never the
inflationary region.  Therefore, livable universes are almost always
created by fluctuations into the ``miraculous" states discussed above.}

\end{center} 

\end{figure}

What then are the alternatives? We may  reject the interpretation of
\des \ based on complementarity. For example, an evolution of the causal
patch based on standard Hamiltonian quantum mechanics may be
wrong. What would replace it is a complete mystery.

Another possibility is an unknown agent intervened in the evolution, and
for reasons of its own restarted the universe in the state of low entropy
characterizing inflation. However, even this does not rid the theory of
the pesky recurrences. Only the first occurrence would evolve in a way
that would be consistent with usual expectations. Subsequently the
recurrences would be extremely unlikely to be consistent, in this sense.
Thus the anthropically acceptable part of spacetime would be dominated by
peculiar, incomprehensible conditions.  This could be avoided if the
system is not ergodic, so that trajectories which pass through inflation
remain in a bounded part of the phase space, and rarely or never enter the
``inconsistent" regions.  This seems very unlikely, and even if true, it
only involves a tiny subset of all the possible trajectories, leaving us
with the still difficult task of explaining why we exist in such an
unusual part of phase space. It is also possible that we are missing some
important feature that picks out, or weights disproportionally, the
recurrences which go through a conventional evolution, beginning with an
inflationary era. However, we have no idea what this feature would be.

We wish to emphasize that the above conclusions appear to be 
the inevitable consequence of the following assumptions:  
\begin{itemize}
\item 
There is a fundamental cosmological constant.
\item
We can apply the ideas of holography and complementarity to de Sitter 
space.
\item
The time evolution operator is unitary, so that phase
space area is conserved.  
\end{itemize}

Perhaps the only reasonable conclusion is that we do not live in 
a world with a true cosmological constant.  

\bigskip

{\centerline {\bf Acknowledgements }}

It is a pleasure to thank A. Albrecht, T. Banks, P. Batra, 			R. Bousso, S. 
Hellerman, A. Linde, J. Lindesay, N. Kaloper 
and S. Shenker for valuable discussions.  This work was supported in part
by NSF grant PHY-9870115 and by the Stanford Institure for Theoretical 
Physics.   
M.K. is the Mellam Family Foundation Graduate Fellow.


\section{Appendix: Poincare Recurrences}

The classical Poincare Recurrence theorem states that the trajectory of
a
phase space point, $(p,q)$, in an isolated, finite system is
quasi-periodic.
If one waits a sufficiently long time, the phase space point will, to
the
desired level of accuracy, continuously recur.  The quantum Poincare
Recurrence theorem is similar and can be stated as follows:
given a system in which all energy eigenvalues are discrete, a state
will return arbitrarily close to its initial value in a finite
amount of time.  It follows immediately that expectation values of
observables
will also return arbitrarily close to their original values.

The classical and quantum recurrence theorems are closely related.
Classically, the phase space point, $(p,q)$ undergoes recurrences while
quantum mechanically, the expectation values $<p>$ and $<q>$ recur.
Additionally, in the limit of a continuous energy spectrum, the quantum
theorem breaks down.  This breakdown corresponds to a classically
unbounded
system and hence the breakdown of the classical theorem as well.

\subsection{Quantum Recurrence Theorem}
The quantum theorem is easy to prove \cite{bocchieri, schulman,
percival} and
we outline the proof below.

Let $\Psi(t)$ be a wave function whose evolution is controlled by
the Schrodinger equation.  For a system with discrete energy
eigenvalues, $\Psi(t)$ can be written in terms of the energy
eigenstates, $u_n$, as

\begin{equation}
\Psi(t) = \sum_{n=0}^{\infty} a_n \exp(-i E_n t) u_n\ .
\end{equation}
We want to show that there exists a time T such that

\begin{equation}
\Vert \Psi(T) - \Psi(0)\Vert  = \left[\sum_{n=0}^{\infty} 2 \ |a_n|^2 (1
-
\cos E_n T)\right]^{1 \over 2} < \epsilon
\end{equation}\label{norm}
for some $\epsilon > 0$.  Since

\begin{equation}
\sum_{n=0}^{\infty} |a_n|^2 = 1\, ,
\end{equation}
there exists an integer N such that

\begin{equation}
\sum_{n=N+1}^{\infty} 2 \ |a_n|^2 (1 - \cos E_n T) \leq
\sum_{n=N+1}^{\infty}
2 \ |a_n|^2 < \epsilon^2.
\end{equation}
We therefore only need to consider the finite sum

\begin{equation}
\sum_{n=0}^{N} 2 \ |a_n|^2 (1 - \cos E_n T) \ .
\end{equation}
But this is an almost periodic function.  The Bohl-Wennberg
theorem \cite{besicovitch} states that there exists a relatively dense
set of
values, $T = T(\delta, M_n)$, such that, for integers $M_n$
and an arbitrarily small $\delta$,

\begin{equation}
| E_n T - M_n 2 \pi| < \delta,
\end{equation}
with $n = 0, 1, 2 \dots N$.  In other words, there is a $T = T(\delta,
M_n)$
such that

\be
T
=  { 2 \pi M_0 + \delta_0 \over E_0}
= { 2 \pi M_1 + \delta_1 \over  E_1 }
= \dots
= { 2 \pi M_N + \delta_N \over  E_N },
\ee
with $|\ \delta_n\ | < \delta$.  This in turn implies

\begin{equation}
\sum_{n=0}^{N} 2 \ |a_n|^2 (1 - \cos E_n T) \ <\ \sum_{n=0}^{N} |a_n|^2
\delta^2 \ < \ \epsilon^2.
\end{equation}

So the wave function of a system with discrete energy eigenvalues will
continuously return arbitrarily close to its initial value.
The above argument, which simply states that a uniformly convergent
Fourier
series is an almost periodic function, can easily be generalized to
show
that expectation values of observables, $<A>$, also undergo recurrences.

\subsection{Correlation Functions}
Since the time one must wait to experience a Poincare recurrence can be
astronomical, the effects on measurements can be small.  These effects,
however, are important when attempting to distinguish large but finite
systems
from infinite ones.  We will consider one such effect below.

Switching to the Heisenberg picture, consider the following correlation
function

\be
F(t) = \ <A(0)A(t)>\ .
\ee
We know that this correlation function will undergo Poincare recurrences
in a
finite system.  Let us approximate the effects on the properly
normalized long
time average of $F(t) F^*(t)$

\be
L = \lim_{T \to \infty} \ {1 \over 2T} \int_{-T}^{T}  \ { |F(t)|^2
\over |F(0)|^2}\ dt\ .
\ee
Assuming that we begin in a state with a density matrix given by

\be
\rho = \exp(-\beta H)\,
\ee
we can insert a complete set of energy eigenstates to obtain

\be
L = {\sum_{mnpq} e^{-\beta(E_n + E_p)}
|A_{mn}|^2 |A_{pq}|^2 \
\delta_{(E_{n} - E_{m} - E_{p} + E_{q})}
\over
\sum_{mnpq} e^{\beta(E_n + E_p)} |A_{mn}|^2 |A_{pq}|^2}.
\ee
We can convert this into an integral by inserting the density of states
$\Omega(E) = \exp(S)$ where $S$ is the entropy of our system.
Since $\delta (E_i) \rightarrow \delta (E)/ \Omega (E)$,  we find

\be
L = {\int dE_1 dE_2 dE_3 \ |A(E_1,E_2)|^2 |A(E_3,E')|^2
\ e^{S_1 +S_2+S_3-\beta(E_1 + E_3)}
\over
\int dE_1 dE_2 dE_3 dE_4  \ |A(E_1,E_2)|^2 |A(E_3,E_4)|^2
\ e^{S_1 +S_2+S_3 + S_4-\beta(E_1 + E_3)}},
\ee
where $E' = E_1 + E_2 - E_3$.  The method of steepest descent
can be used to show

\be
L \sim \exp(-S).
\ee
Although $L$ may be incredibly small for sufficiently large
systems, it will not vanish since the entropy of a finite
system is finite.  It will vanish, however, for systems that
are infinite.  This is one example of a result which differs
for finite and infinite systems.


\begin{thebibliography}{999}
\bibitem{tom}
Complementarity in the context of deSitter Space is also
discussed by T.Banks and W.Fischler in ``M-theory Observables for
Cosmological
Space-times.''
\end{thebibliography}
\end{document}